\documentclass[apsrev4,pra,superscriptaddress,notitlepage,nofootinbib]{revtex4-1} 
\usepackage[colorlinks=true,allcolors=blue]{hyperref}
\usepackage{amsmath,amsfonts,amssymb}
\usepackage{graphicx,xcolor}
\usepackage[version=4]{mhchem}
\begin{document} 
\title{Application of nanodiamond-polymer composite holographic gratings in a very cold neutron interferometer}
\author{Sonja Falmbigl}
\author{Roxana H. Ackermann}
\affiliation{Faculty of Physics, University of Vienna, Wien, Austria}
\author{Elhoucine Hadden}\thanks{now at: Department of Physics and Astronomy, Uppsala University, Sweden}
\affiliation{Institut Laue-Langevin, Grenoble, France}
\author{Hanno Filter-Pieler}
\author{Tobias Jenke}
\author{J\"urgen Klepp}
\affiliation{Faculty of Physics, University of Vienna, Wien, Austria}
\author{Christian Pruner}
\affiliation{Faculty of Natural and Life Sciences, Department of Chemistry and Physics of Materials, University of Salzburg, Austria}
\author{Yasuo Tomita}
\affiliation{Department of Engineering Science, University of Electro-Communications, Tokyo, Japan}
\author{Martin Fally}
\thanks{send correspondence to M.F.: martin.fally@univie.ac.at}
\affiliation{Faculty of Physics, University of Vienna, Wien, Austria}

\hypersetup{pdfauthor={Falmbigl, Ackermann, Hadden, Filter-Pieler, Jenke, Klepp, Pruner, Tomita, Fally},pdftitle={Application of nanodiamond-polymer composite holographic gratings in a very cold neutron interferometer},pdfborder={0 0 0},urlcolor=blue}
\newcommand{\bcr}{[b_c\Delta\rho]_1}
\definecolor{ured}{RGB}{167,28,73}
\definecolor{ublue}{RGB}{0,99,166}
\definecolor{ugreen}{RGB}{148,193,84}
\definecolor{ugray}{RGB}{102,102,102}
\definecolor{uorange}{RGB}{221,72,20}
\definecolor{uyellow}{RGB}{246,168,0}
\definecolor{umint}{RGB}{17,137,122}
\newcommand{\tcr}[1]{\textcolor{ured}{#1}}
\newcommand{\tcb}[1]{\textcolor{ublue}{#1}}
\newcommand{\yco}[1]{\textcolor{black}{#1}}
\newcommand{\eco}[1]{\textcolor{black}{#1}}
\newcommand{\hco}[1]{\textcolor{black}{#1}}

\graphicspath{{/home/fallym4/MyConferences/25/SPIE_Prague25/LaTeX/figs/}}

\begin{abstract}
\eco{In recent decades, photosensitive materials have been used for the development of optical devices not only for light, but also for cold and very cold neutrons. 
We show that holographically recorded gratings in nanodiamond-polymer composites \hco{(nDPC)} form ideal diffraction elements for very cold neutrons. Their advantage of high diffraction efficiency, combined with low angular selectivity as a two-port beam splitter, meets the necessary conditions for application in a very cold neutron interferometer.} 
We provide an overview of the latest achievements in the construction of such a triple Laue interferometer. A first operational test of the interferometer is planned immediately after this conference in May 2025.
\end{abstract}

\maketitle
\section{INTRODUCTION}
\label{sec:intro}  
Half a century ago the first neutron interferometer with widely separated coherent beams based on Bragg diffraction from a perfect crystal boosted neutron optical and quantum mechanical experiments with (thermal) neutrons \cite{Rauch-pla74}. This kind of interferometer is akin to a Mach-Zehnder interferometer in light optics by replacing the beamsplitters as well as the mirrors with thick diffractive crystals in transmission geometry. While three equidistant and identical silicon crystal slabs, in the so called triple-Laue arrangement (LLL) with perfect alignment of the corresponding diffraction vectors, have been widely used for thermal neutrons (for an overview see Ref. [\citenum{Klepp-ptep14,Rauch-15}]), they cannot be used for neutrons with wavelengths exceeding twice the lattice spacing. To even enlarge the size, and thus the sensitivity, of the interferometer a split-crystal setup was successfully tested\cite{Lemmel-jac22}.
An extension beyond the thermal energies was established by using phase gratings \cite{Gruber-pla89,Eder-nima89,Zouw-nima00} for very cold neutrons (VCN). Surface-relief gratings \cite{Ioffe-pla85}, multilayer mirrors \cite{Funahashi-pra96} or artificial, holographically prepared thick gratings in photopolymers inspired by their thermal neutron counterpart were employed for the cold neutron regime \cite{Schellhorn-pb97,Pruner-nima06}. 

The sensitivity of an interferometer depends on the phase difference $\Delta\Phi$ accumulated in either path \cite{Rauch-15}. For an LLL-type interferometer we aim at well separated paths which requires a long distance between each of the diffraction gratings. Here, we discuss our recent attempt to set up a new interferometer for the VCN regime with a total length of $L=1500\,$mm, gratings with a spacing of $\Lambda\approx 500\,$nm and thus \eco{an estimated beam separation at the second grating} of about $7.5\,$mm for VCN with a wavelength of $5\,$nm. Given a neutron beam with a certain characteristics there remain a vast number of severe challenges to assemble an LLL-interferometer for VCN (see Ref. \citenum{Zouw-PhD00}). Among them the most important ones are (1) the fabrication of appropriate gratings, and (2) their ultra-accurate alignment. 

Our approach is to control the rotational alignment during the fabrication process of the gratings with respect to a precise \hco{reference granite plate}, ensuring roll, pitch and yaw of each grating to be within about a $\mu\text{rad}$. This is sufficiently accurate according to detailed \hco{studies} \cite{Schellhorn-PhD98,Breer-M95} and also to our experience with an interferometer of similar type but shorter length ($L=300\,$mm). \cite{Pruner-PhD04,Pruner-nima06} A general overview on the requirements on artificial gratings for a VCN interferometer and the optimal choice of materials can be found in Ref. \citenum{Fally-spie24}.

\section{PREPARATION OF GRATINGS AND LIGHT OPTICAL CHARACTERISTICS}
Different groups of holographic materials were tested over the last 20 years for their applicability as efficient beamsplitters in cold neutron diffraction and VCN diffraction experiments \cite{Fally-apb02,Klepp-m12,Fally-spie24}. Finally, nanodiamond polymer composites (nDPC) turned out to be the best choice for gratings used in a VCN interferometer \eco{\cite{Tomita-spie20,Fally-spie20,Hadden-spie22,Hadden-apl24}}. To prepare the material we followed the formulation given in Ref. \citenum{Tomita-pra20} except for replacing the photoinitiator with Rose Bengal (RB), a dye that is photosensitive in the green optical region with a maximum of absorbance around 565 nm. A series consisting of more than hundred samples was prepared to optimize the grating quality with respect to recording \hco{light} intensity, sample and grating thickness, pre- and post-exposure, and the concentration of the RB, respectively. 

\begin{figure}[ht]
   \begin{center}
   \begin{tabular}{cc} 
    \includegraphics[width=\columnwidth/2]{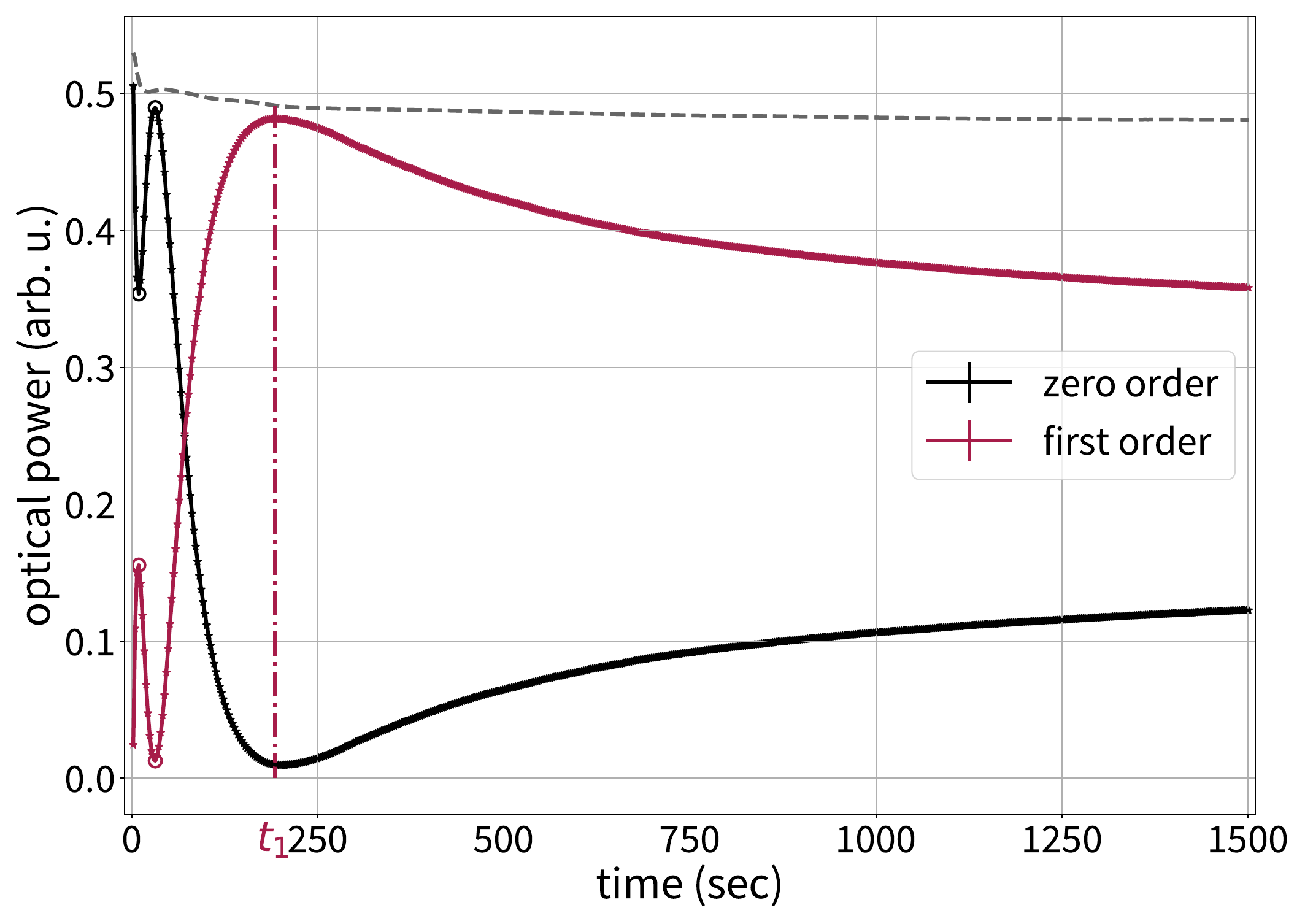}&\includegraphics[width=\columnwidth/2]{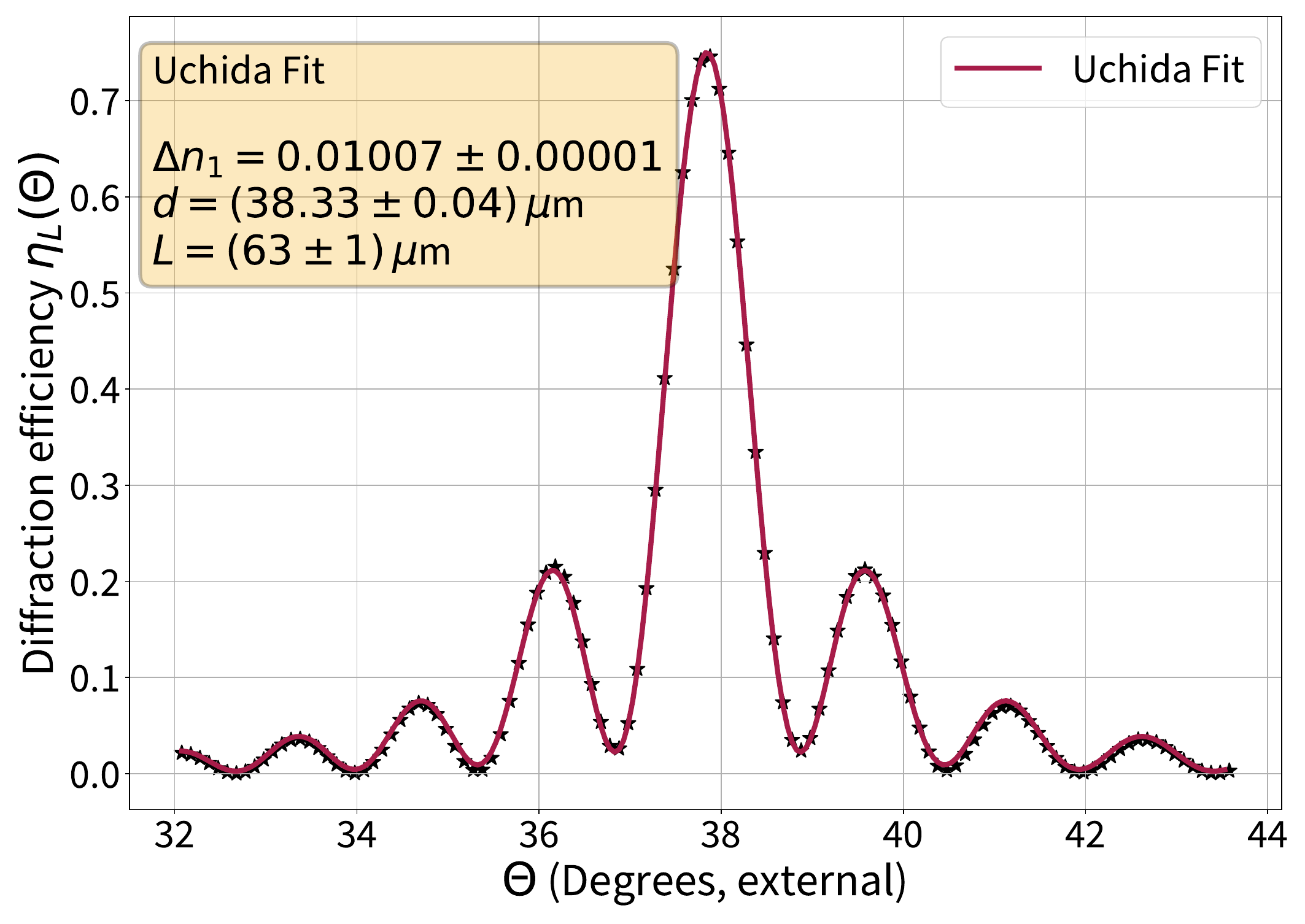}
   \end{tabular}
   \end{center}
   \caption[light1Lred] 
   { \label{fig:1} 
\textit{Left panel}: Buildup dynamics of the grating's first and zero order diffraction power (raw data) at Bragg matching condition as a function of recording time. The gray dashed line represents the total optical power. \textit{Right panel}: Angular dependence of the diffraction efficiency $\eta_\text{\tiny L}(\Theta)$ after saturation. The monitioring wavelength was $\lambda_\text{\tiny L,red}=632.8~\textrm{nm}$ to which the holographic material nDPC is insensitive.}
   \end{figure}

\newcommand{\lri}{n_\text{\tiny L}}
Fabrication of (volume) gratings for the VCN interferometer is performed as usual by holographic recording of two plane, s-polarized, coherent light waves which intersect at \eco{the nDPC sample position ($2\Theta\approx 61.2^\circ$)}. The \eco{used} recording wavelength is $\lambda_\text{\tiny L}=514\,$nm, the intensity at the position of the sample $I=200\,\text{mW/cm}^2$. This results in a one-dimensional periodic modulation, w.l.o.g. along the $x$-direction, of the refractive index for light. Light optical diffraction characteristics for the zero and first diffraction orders at Bragg matching condition during recording and after saturation and postcuring are shown in Fig. \ref{fig:1}. The wavelength used for monitoring was $\lambda_\text{\tiny L,red}=632.8\,$nm at which the nDPC syrup is insensitive.

The temporal behaviour of grating formation has an intial phase with a rather fast increase in diffracted power, passing a maximum (at $t\approx 10\,$sec) followed by a minimum (at $t\approx 30\,$sec). 
\yco{
This particular behavior of the grating buildup, as seen in Fig.\,\ref{fig:1}, can be attributed to the emergence of a secondary grating which is due to the mutual diffusion of nanodiamonds and monomer, followed by the fast buildup and decay of a polymer grating (before monomer diffusion under holographic exposure) at high recording intensity ($200\,\text{mW/cm}^2$). The initial grating is in phase with the intensity-interference pattern, while the fundamental component of the secondary one experiences a $\pi$-phase shift because the refractive index of  nanodiamonds is higher than that of the polymer formed.}

Thereafter, the desired grating growth process ramps up and reaches the saturated value $\Delta n_\text{sat}> 0.01$ at a photorefractive sensitivity $S_{\eta2}=20\,$cm/J, which is defined as the slope of $\sqrt{\eta(t)}$ during the initial rise time, divided by intensity and thickness \cite{Hatano-07}. Note, that at $t_1$ the grating becomes overmodulated for light. After recording and postcuring with a (white) LED lamp overnight, the angular dependence of the diffraction efficiency was measured. The result is shown in Fig. \ref{fig:1} together with a fit, accounting for a tentative exponential decay of the first order refractive-index modulation $\Delta n_1(z)$ along the sample depth. It is notable, that in contrast to previously prepared nDPC gratings \cite{Hadden-apl24} apodization, i.e., the lifting of side minima, is negligible. 
\yco{This is expressed by an effective grating thickness $L$ (i.e., the $1/e$ value of the decaying grating along the thickness direction) which is much larger than the mechanical grating thickness $d$}.

\section{NEUTRON OPTICAL CHARACTERISTICS}
\eco{The recorded one-dimensional periodic structure of the formed gratings applies not only to light but also to neutrons for which it induces a modulation of the neutron-refractive index (or equivalently, the neutron optical potential)}. The latter can be written as $n(x)=n_0+2|n_1|\cos(2\pi x/\Lambda+\varphi_1)+2|n_2|\cos(4\pi x/\Lambda+\varphi_2)+\ldots$. 
For a majority of the materials under investigation the relative phases are $\varphi_j\in\{0,\pi\}$, thus $\Delta n_j=\pm 2|n_j|, \varphi\equiv 0$. This is also the case for nDPC, \eco{with $\varphi_1=\pi$ being the phase between the intensity pattern and the fundamental, first order Fourier component of the neutron-refractive index.}
The Fourier-coefficients $n_j$ are related to the essential neutron optical parameter, the coherent scattering length density (SLD) $b_c\rho$ by:
\begin{equation}\label{eq:nRI}
n_0=\frac{\lambda^2[b_c\rho]}{2\pi}; \quad 2n_j=\Delta n_j=\frac{\lambda^2}{2\pi}\Delta[b_c\rho]_j,\, j>0.
\end{equation}
Here we denote the Fourier coefficients of the light-induced changes of the SLD by $\Delta[b_c\rho]_j$. 

The major advantage of \eco{an} nDPC grating over other excellent light optical materials is their applicability for neutron diffraction. They have exceptionally large SLD modulation values $\Delta[b_c\rho]_j$ which, together with the thickness $d$ determine the \eco{maximum achievable diffraction efficiency at a given probe wavelength}. Another important aspect is their negligible incoherent scattering.

\begin{figure}[ht]
   \begin{center}
   \begin{tabular}{cc} 
    \includegraphics[width=\columnwidth/2]{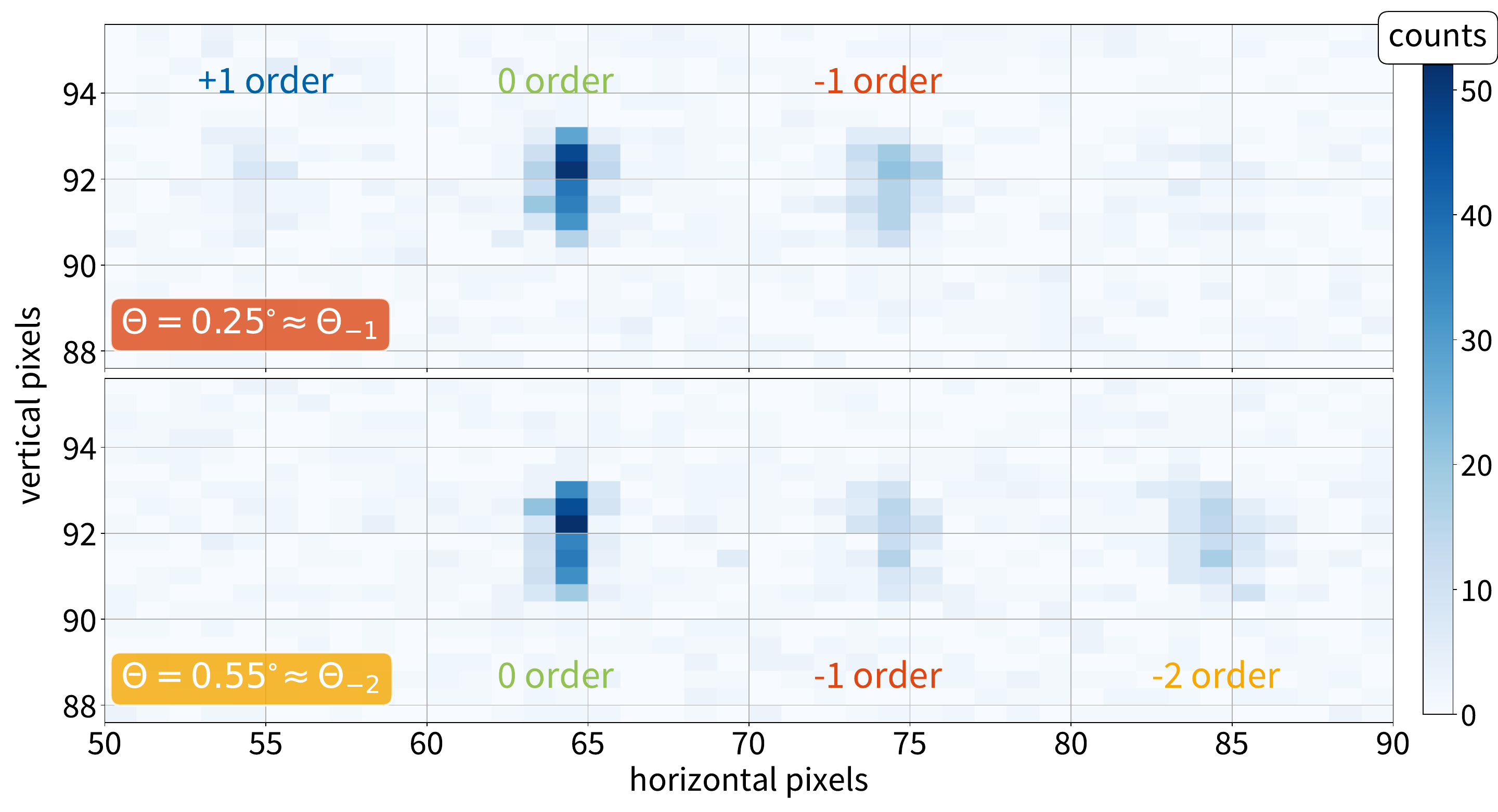}
    &
    \includegraphics[width=\columnwidth/2]{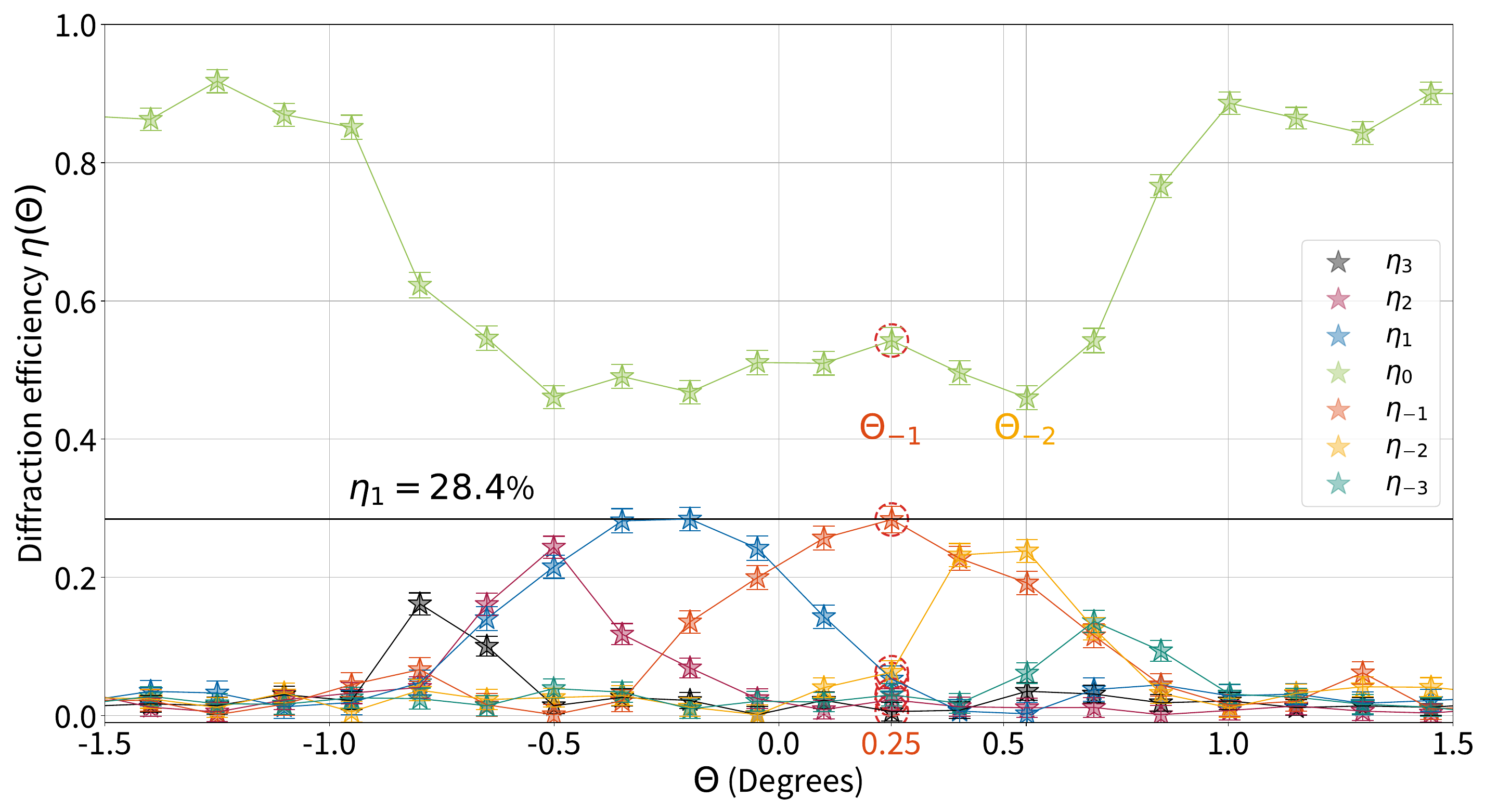}
   \end{tabular}
   \end{center}
   \caption{ \label{fig:neutron} 
\textit{Left panel}: Part of the neutron count map on the detector matrix for two different angular positions (false colors, linear scale). The upper image shows the intensity map near the Bragg position $\Theta_{-1}$ for the $-1^\text{st}$ order, the lower one at $\Theta_{-2}$ for the $-2^\text{nd}$ order.\\
\textit{Right panel}: Angular dependence of the diffraction efficiency $\eta(\Theta)$ for VCN. Diffraction orders between $m=-3,\ldots,+3$ with significant efficiency were detected. The circles mark $\eta_m$ at the Bragg position $\Theta_{-1}$ for which the neutron count map is shown in the upper left panel.}
   \end{figure}

For simplicity and as a rough estimation of diffraction efficiency\footnote{In neutron optics and interferometry usually the diffraction efficiency is termed \textit{reflectivity}.} let us assume that diffraction takes place in the Bragg regime, in which only two waves with significant amplitudes are propagating \cite{Gaylord-ao81}. Then the diffraction efficiency for Bragg-matched order $m$  is approximately
\begin{equation}\label{eq:nDE}
\eta_m=\sin^2\left(\frac{1}{2}d\lambda\Delta[b_c\rho]_m\right).
\end{equation}
However, for realistic cases in VCN diffraction experiments, depending on $d, \lambda$ and $\Delta[b_c\rho]_j$, diffraction takes place in a multiwave-coupling regime \cite{Klepp-jpcs16,Fally-oex21,Hadden-PhD24} and $\eta_m$ is also determined by other Fourier coefficients $\Delta[b_c\rho]_j$ with $j\not=m$. Together with the unavoidable wavelength distribution of the VCN beam this will lead to (usually) lower values than estimated by Eq. (\ref{eq:nDE}). Furthermore, \eco{in most practical cases}, the magnitude of the Fourier coefficients decreases with increasing order $j$. Therefore, we intend to use only the first diffraction order with efficiency $\eta_1$ in the interferometer. The second diffraction order with efficiency $\eta_2$ might still be quite powerful as can be seen in Fig. \ref{fig:neutron}. However, due to dephasing, i.e. deviation from perfect Bragg matching, the second order diffraction can be suppressed provided that $\Lambda/d\approx\Theta_2\approx\lambda/\Lambda$. This is the case for choosing a thickness $d\approx\Lambda^2/\lambda$. For the VCN beamline with a (mean) wavelength of $5\,$nm and a grating period of $500\,$nm the \yco{optimal} thickness thus is $d=50\mu\text{m}$. 
\yco{As can be seen from the angular dependence of the diffraction efficiencies $\eta_m$ in Fig.\,\ref{fig:neutron}, even a grating with a thickness of $d\approx 40\,\mu$m is close enough to suppress higher orders. An accurate calculation requires the use of a theory considering more than two diffraction orders and the nonsinusoidal refractive index profile, e.g., a coupled wave analysis \cite{Moharam-josaa95,Klepp-jpcs16}}.

The properties of the nDPC gratings were examined with VCN at the PF2/VCN beamline of the Institut Laue-Langevin ILL), Grenoble, France. We arranged a small angle diffraction experiment by collimating the VCN beam (mean wavelength: $\lambda=4.6\,$ nm) employing two rectangular slits with 1 mm width at a distance of 1220 mm resulting in a horizontal divergence of less than 2 mrad. The grating is placed on a rotation stage close to the end of the collimation. Finally, the 2D detector with $2\times 2 \text{mm}^2$ pixel size is located 2280 mm downstream and allows to observe spots of all relevant diffraction orders ($m=\pm 3, \pm 2, \pm 1, 0$). To prevent scattering of the VCN by air, the beam path is shielded with He-filled tubes. The diffraction efficiency of order $m$ is defined as the neutron counts diffracted to order $m$ divided by the sum of all counts: $\eta_m=I_m/\sum I_m$. To examine the properties of the grating we measured the counts of diffracted neutrons at a range of angular values $\Theta$; data are made available by the ILL \cite{Klepp-ddoi24}. \hco{The subtraction of the background follows one or the other method presented in Ref.\,} \citenum{Hadden-PhD24}. The angular dependence of the resulting diffraction efficiencies is shown in Fig. \ref{fig:neutron} together with neutron count maps for two angular positions, either near the minus first or second order Bragg angle $\Theta_{-1;-2}$, respectively.

\begin{figure}[ht]
   \begin{center}
   \begin{tabular}{c} 
     \includegraphics[width=\columnwidth]{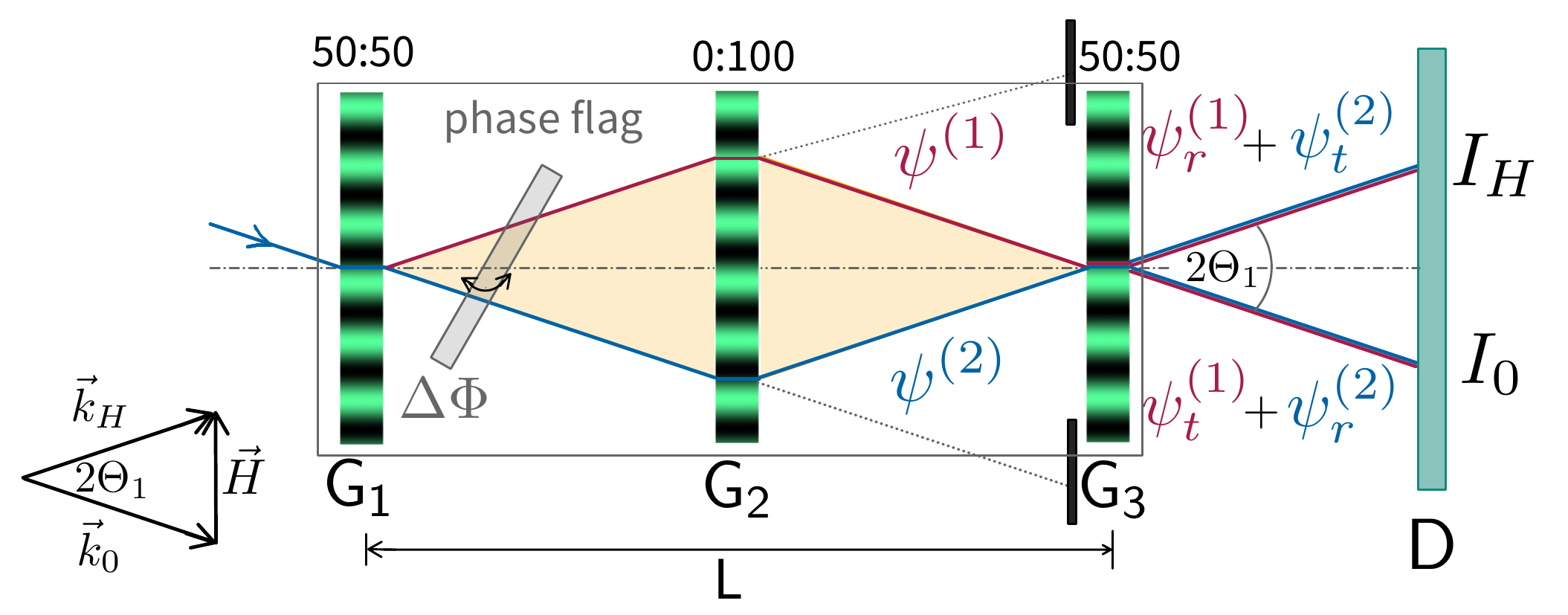}
   \end{tabular}
   \end{center}
   \caption{ \label{fig:IF} 
   Sketch of the VCN interferometer and beam paths therein. G1--G3 are the gratings, D denotes the detector and $\tcr{\psi^{(1)}},\tcb{\psi^{(2)}}$ are the wavefunctions for the corresponding path at G3. The subscripts $r,t$ denote the wavefunctions reflected or transmitted from G3, respectively. The detected intensities are $I_0=|\tcr{\psi_t^{(1)}}+\tcb{\psi_r^{(2)}}|^2$ and $I_H=|\tcr{\psi_r^{(1)}}+\tcb{\psi_t^{(2)}}|^2$. Lower left drawing: the Laue condition, coining $0,H$ beam directions.
}
   \end{figure}

The nDPC grating shows excellent diffraction properties. For the first orders we find an efficiency of about 28\% at the corresponding Bragg angles $\Theta_1$ at which the second orders exhibt a minimum. This is important for the neutron flux economy when multiwave-coupling occurs. Note, that the measured diffraction efficiency is even higher than for gratings reported in a recent publication \cite{Hadden-apl24}, as therein (1) the wavelength was longer, see Eq. \ref{eq:nDE}, or/and (2) the grating was tilted by $\zeta\approx 70^\circ$ about the its symmetry axis which leads to more than double effective thickness. While this latter approach is very useful, it is not applicable for an LLL-interferometer. We also prepared thicker gratings up to $d\approx 60\,\mu$m with diffraction efficiencies of the untilted grating higher than $50\%$. However, when it comes to VCN one drawback at this thickness is the higher angular selectivity. Keeping in mind that the VCN beam has a rather broad wavelength distribution, part of the spectra will not be sufficiently diffracted. That is why \eco{slightly thinner gratings are} \hco{chosen} for the interferometer.
After having fabricated three gratings, their mutual alignment is key to success. Perfectly aligned gratings with a flux maximized LLL interferometer and simultaneous highest contrast would be established for grating G1 and G3 a 50:50 beamsplitter, and G2 a mirror, respectively. By impressing a phase difference of $\Delta\Phi$ between the two paths, maximum visibility  $v=(I_\text{max}-I_\text{min})/(I_\text{max}+I_\text{min})=1$ for both pairs of the interfering beams with intensities $I_H, I_0$, respectively, are obtained (see sketch of beampaths in Fig. \ref{fig:IF}).
For $\eta_1=\eta_3=\eta$ but $0<\eta_2<1$ the visibility in the interfering beams travelling along $0$-direction remains still perfect ($v_0=1$), whereas the pair's visibility travelling along the $H$-direction is $v_H(\eta)=2\eta(1-\eta)/[1-2\eta(1-\eta)]$. This is a realistic case. By shielding the beams forward-diffracted from G2, high visibility is possible, however, the experiment suffers from loss of neutron flux onto the detector.
\section{ALIGNMENT OF THE GRATINGS}
Alignment of the three gratings is utmost delicate and requires ultra-accurate adjustment. Any deviation from perfect alignment and/or variation of the grating properties lead to a loss of contrast (visibility). Among them are: deviation of the grating period for G1-G3, bent gratings, curved sample surface and their roughness; furthermore a deviation from equidistance between G1/G2 and G2/G3, pitch, yaw and roll angles. While all of them add up, the latter is the most critical one. To ensure less than 10\% loss of contrast, the roll angle accuracy must not exceed a few microradians.
Detailed discussions and estimated accuracy requirments can be found in \cite{Drabkin-nima89,Breer-M95,Schellhorn-PhD98,Zouw-PhD00}.

In a previously constructed interferometer for cold neutrons we mastered the alignment requirements by mounting the three samples of photosensitive deuterated poly(methyl)methacrylate equidistantly placed on a \hco{rigid} breadboard, making it portable and \hco{deployable} at different beamlines. To prepare the gratings each of the slabs was translated along the interferometer's symmetry axis into a standing light intensity pattern (two-wave mixing). After the translation and before exposing the next slab, the accuracy of the translation was detected by an autocollimation instrument (pitch, yaw) and a polarization optical setup sensitive to roll angle deviations \cite{Pruner-02,Pruner-PhD04,Pruner-nima06}. This allowed us to successfully demonstrate the operation of the interferometer.

\begin{figure}[ht]
   \begin{center}
   \begin{tabular}{c} 
\includegraphics[height=4.5cm]{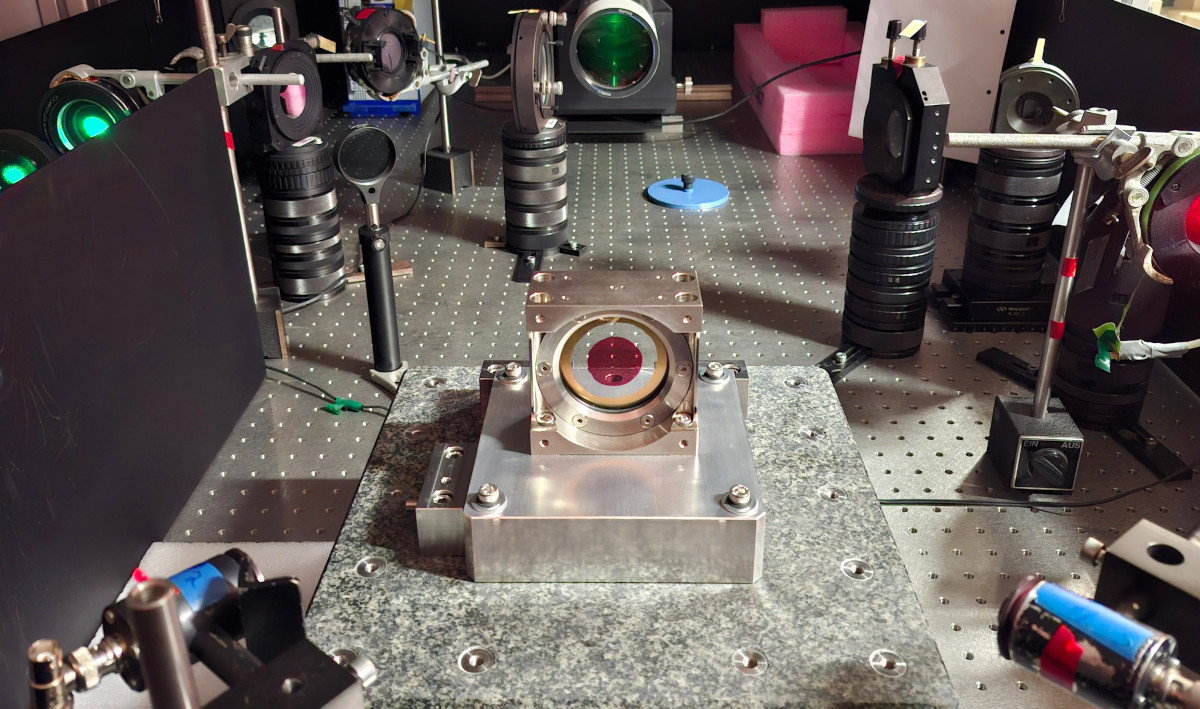}
\frame{\includegraphics[height=4.5cm]{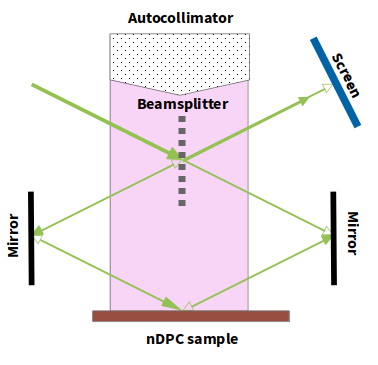}}
\includegraphics[height=4.5cm]{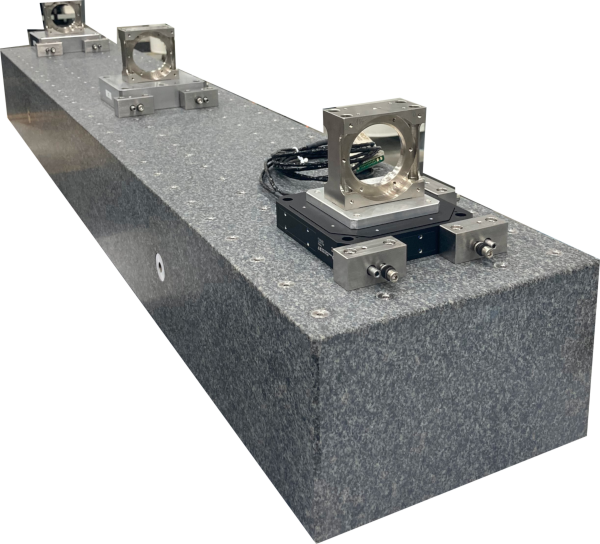}
   \end{tabular}
   \end{center}
   \caption{ \label{fig:VCN} 
\textit{Left panel}: Picture of the optical recording. \textit{Center panel}: Schematic of the adjustment setup, top view. Autocollimator and bisectrix of the recording angle are accurately in line within $\mu$rad. \textit{Right panel}: The GB with roughly aligned sample holders. \copyright Fa.\,Johann Fischer, Aschaffenburg}
   \end{figure}
To increase the size of the interferometer to $L=1.5\,$m we adopted a different approach. Removeable sample holders are placed on a super-planar plate of aluminum or a 6-axis nanopositioning stage for G3, respectively, which were adjusted on an extremely flat granite bench (GB) ($300\times 1760\,\text{mm}^2$) by \textit{Fa. Johann Fischer, Aschaffenburg}. The maximum deviation over the total area is less than half a micrometer; this defines the accuracy of roll and pitch angles. A shorter but otherwise identical twin of the granite bench ($300\times 300\,\text{mm}^2$) is located on an optical table at the optical lab with a standard holographic two-wave mixing setup. 
The nDPC sample in its holder is placed on the granite and its surface-normal accurately aligned to form the bisectrix of the two holographic recording beams thereby defining zero yaw angle. Additionally we bring an autocollimation system (\textit{Trioptics}) into line to exactly conincide with the bisectrix, which later serves as the reference when adjusting the gratings on the large bench at the neutron facility. A schematic and picture of the recording and adjustment setup are shown in Fig. \ref{fig:VCN}.
After recording of the gratings in nDPC they are transferred to PF2/VCN instrument and installed on the large GB, employing the autocollimation system for each of them to control and correct the yaw angle; pitch and roll angles are expected to fit. A picture of the GB and the sample holders in place during the calibration at \textit{Fa. Johann Fischer} is shown in Fig. \ref{fig:VCN}. The GB will be placed on a support, which can be moved on air cushions for positioning the VCN interferometer with respect to the neutron beam. A vacuum chamber will be designed to cover the interferometer and avoid scattering from air. The assembly and a first operational test are scheduled for May 2025.
\section{SUMMARY}
We discussed the setup of an interferometer for VCN based on artificial gratings in the LLL-geometry. Gratings are holographically recorded in nDPC with desired grating period, thickness, and to a certain extent diffraction efficiency. The accurate mutual alignment of the gratings (within $\mu$rad in pitch, yaw and roll angles) in a VCN interferometer is accounted for by using a super-flat granite bench and an autocollimator. After successful setup results of a first operational test will be reported elsewhere.

\acknowledgments 
Yasuo Tomita acknowledges a visiting professorship of the University of Vienna. \hco{Our special thanks go to Thomas Brenner who has provided us with crucial technical support on the very cold neutron beamline PF2/VCN over two decades}.
This research was funded in part by the Austrian Research Promotion Agency (FFG), Quantum-Austria NextPi, grant number FO999896034 and the European Union:
”NextGenerationEU”.


\begin{thebibliography}{10}

\bibitem{Rauch-pla74}
H.~Rauch, W.~Treimer, and U.~Bonse, ``Test of a single crystal neutron
  interferometer,'' {\em Phys. Lett. A} {\bf 47}, p.~369, 1974.
\doi{10.1016/0375-9601(74)90132-7}

\bibitem{Klepp-ptep14}
J.~Klepp, S.~Sponar, and Y.~Hasegawa, ``Fundamental phenomena of quantum
  mechanics explored with neutron interferometers,'' {\em Prog. Theor. Exp.
  Phys.} {\bf 2014}, August 2014.
\newblock \doi{10.1093/ptep/ptu085}.

\bibitem{Rauch-15}
H.~Rauch and S.~A. Werner, {\em Neutron interferometry}, Oxford University
  Press, New York--Oxford, 2nd~ed., 2015.
\newblock \doi{10.1093/acprof:oso/9780198712510.001.0001}.

\bibitem{Lemmel-jac22}
H.~Lemmel, M.~Jentschel, H.~Abele, F.~Lafont, B.~Guerard, C.~P. Sasso, G.~Mana,
  and E.~Massa, ``Neutron interference from a split-crystal interferometer,''
  {\em J. Appl. Crystallogr.} {\bf 55}, pp.~870--875, July 2022.
\newblock \doi{10.1107/s1600576722006082}.

\bibitem{Gruber-pla89}
M.~Gruber, K.~Eder, A.~Zeilinger, R.~G{\"a}hler, and W.~Mampe, ``A
  phase-grating interferometer for very cold neutrons,'' {\em Phys. Lett. A}
  {\bf 140}, p.~363, 1989.
\doi{10.1016/0375-9601(89)90068-6}.

\bibitem{Eder-nima89}
K.~Eder, M.~Gruber, A.~Zeilinger, R.~G{\"a}hler, W.~Mampe, and W.~Drexel, ``The
  new very-cold-neutron optics facility at {ILL},'' {\em Nucl. Instrum. Meth.
  A} {\bf 284}, pp.~171--175, November 1989.
\doi{10.1016/0168-9002(89)90273-8}.

\bibitem{Zouw-nima00}
G.~van~der Zouw, M.~Weber, J.~Felber, R.~G{\"a}hler, P.~Geltenbort, and
  A.~Zeilinger, ``{Aharonov-Bohm} and gravity experiments with the
  very-cold-neutron interferometer,'' {\em Nucl. Instrum. Meth. A} {\bf 440},
  pp.~568--74, 2000.
\doi{10.1016/s0168-9002(99)01038-4}.

\bibitem{Ioffe-pla85}
A.~Ioffe, V.~Zabiyakin, and G.~Drabkin, ``Test of a diffraction grating neutron
  interferometer,'' {\em Phys. Lett. A} {\bf 111}, pp.~373--375, September
  1985.
\newblock \doi{10.1016/0375-9601(85)90373-1}.

\bibitem{Funahashi-pra96}
H.~Funahashi, T.~Ebisawa, T.~Haseyama, M.~Hino, A.~Masaike, Y.~Otake,
  T.~Tabaru, and S.~Tasaki, ``Interferometer for cold neutrons using multilayer
  mirrors,'' {\em Phys. Rev. A} {\bf 54}, p.~649, 1996.
\newblock \doi{10.1103/physreva.54.649}.
  
\bibitem{Schellhorn-pb97}
U.~Schellhorn, R.~A. Rupp, S.~Breer, and R.~P. May, ``The first neutron
  interferometer built of holographic gratings,'' {\em Physica B} {\bf
  234-236}, pp.~1068--1070, 1997.
\newblock \doi{10.1016/S0921-4526(97)00015-X}.

\bibitem{Pruner-nima06}
C.~Pruner, M.~Fally, R.~A. Rupp, R.~P. May, and J.~Vollbrandt, ``Interferometer
  for cold neutrons,'' {\em Nucl. Instrum. Meth. A} {\bf 560}, pp.~598--605, 5
  2006.
\newblock \doi{10.1016/j.nima.2005.12.240}.

\bibitem{Zouw-PhD00}
G.~van~der Zouw, {\em Gravitational and {Aharonov-Bohm} phases in Neutron
  Interferometry}.
\newblock PhD thesis, Universit\"at Wien, Austria, 2000.

\bibitem{Schellhorn-PhD98}
U.~Schellhorn, {\em {Anwendung holographischer Gitter in der Neutronenoptik}}.
\newblock PhD thesis, {Universit\"at Osnabr\"uck}, Germany, 1998.
\newblock (in German).

\bibitem{Breer-M95}
S.~Breer, ``{Konstruktion und Aufbau eines {LLL}-Interferometers},'' Master's
  thesis, {Universit\"at Osnabr\"uck}, Germany, 1995.
\newblock (in German).

\bibitem{Pruner-PhD04}
C.~Pruner, {\em {Ein Interferometer f\"ur kalte Neutronen}}.
\newblock PhD thesis, {Universit\"{a}t Wien}, Austria, 2004.
\newblock (in German).

\bibitem{Fally-spie24}
M.~Fally, J.~Klepp, C.~Pruner, E.~Hadden, A.~Bianco, J.~Kohlbrecher, H.~Filter,
  T.~Jenke, and Y.~Tomita, ``Photosensitive materials for neutron optics,'' in
  {\em Photosensitive Materials and their Applications III},  R.~R. McLeod,
  Y.~Tomita, and I.~Pascual~Villalobos, eds., vol.~13015, p.~130150O, SPIE,
  June 2024.
\newblock \doi{10.1117/12.3022432}.

\bibitem{Fally-apb02}
M.~Fally, ``The photo-neutronrefractive effect,'' {\em Appl. Phys. B} {\bf 75},
  pp.~405--426, October 2002.
\doi{10.1007/s00340-002-1035-0}.

\bibitem{Klepp-m12}
J.~Klepp, C.~Pruner, Y.~Tomita, P.~Geltenbort, J.~Kohlbrecher, and M.~Fally,
  ``Holographic gratings for slow-neutron optics,'' {\em Materials} {\bf 5},
  p.~2788, 2012.
\newblock \doi{10.3390/ma5122788}.

\bibitem{Tomita-spie20}
Y.~Tomita, A.~Kageyama, Y.~Iso, K.~Umemoto, M.~Liu, J.~Klepp, C.~Pruner,
  T.~Jenke, P.~Geltenbort, and M.~Fally, ``Nanodiamond-polymer composite
  gratings as diffractive optical elements for light and neutrons {I}: their
  fabrication and light optical diffraction properties,'' in {\em
  Photosensitive Materials and their Applications},  R.~R. McLeod, I.~P.
  Villalobos, Y.~Tomita, and J.~T. Sheridan, eds., {\em SPIE Proc. Ser.} {\bf
  11367}, p.~113670M, SPIE, 2020.
\doi{10.1117/12.2555651}.

\bibitem{Fally-spie20}
M.~Fally, J.~Klepp, C.~Pruner, T.~Jenke, P.~Geltenbort, A.~Kageyama, Y.~Iso,
  K.~Umemoto, M.~Liu, and Y.~Tomita, ``Nanodiamond-polymer composite gratings
  as diffractive optical elements for light and neutrons {II}: neutron optical
  diffraction properties,'' in {\em Photosensitive Materials and their
  Applications},  R.~R. McLeod, I.~P. Villalobos, Y.~Tomita, and J.~T.
  Sheridan, eds., {\em Proc. SPIE} {\bf 11367}, p.~113670N, SPIE, 2020.
\doi{10.1117/12.2555474}.

\bibitem{Hadden-spie22}
E.~Hadden, Y.~Iso, A.~Kume, K.~Umemoto, T.~Jenke, M.~Fally, J.~Klepp, and
  Y.~Tomita, ``Nanodiamond-based nanoparticle-polymer composite gratings with
  extremely large neutron refractive index modulation,'' in {\em Photosensitive
  Materials and their Applications II},  R.~R. McLeod, I.~P. Villalobos,
  Y.~Tomita, and J.~T. Sheridan, eds., {\em Proc. SPIE} {\bf 12151},
  pp.~1215109--1, SPIE, 2022.
\doi{10.1117/12.2623661}.

\bibitem{Hadden-apl24}
E.~Hadden, M.~Fally, Y.~Iso, T.~Jenke, J.~Klepp, A.~Kume, K.~Umemoto, and
  Y.~Tomita, ``{Holographic nanodiamond–polymer composite grating with
  unprecedented slow-neutron refractive index modulation amplitude},'' {\em
  Appl. Phys. Lett.} {\bf 124}(7), p.~071901, 2024.
\doi{10.1063/5.0186753}.

\bibitem{Tomita-pra20}
Y.~Tomita, A.~Kageyama, Y.~Iso, K.~Umemoto, A.~Kume, M.~Liu, C.~Pruner,
  T.~Jenke, S.~Roccia, P.~Geltenbort, M.~Fally, and J.~Klepp, ``Fabrication of
  nanodiamond-dispersed composite holographic gratings and their light and
  slow-neutron diffraction properties,'' {\em Phys. Rev. Appl.} {\bf 14},
  p.~{044056}, 2020.
\newblock \doi{10.1103/PhysRevApplied.14.044056}.
  
\bibitem{Hatano-07}
H.~Hatano, Y.~Liu, and K.~Kitamura, {\em Photorefractive Materials and Their
  Applications 2: Materials}, vol.~114 of {\em Springer Series in Optical
  Sciences}, ch.~Growth and PhotorefractiveProperties of Stoichiometric
  \ce{LiNbO3} and \ce{LiTaO3}, pp.~127--164.
\newblock Springer, New York, 2007.
\newblock \doi{10.1007/0-387-34081-5}.

\bibitem{Gaylord-ao81}
T.~K. Gaylord and M.~G. Moharam, ``Thin and thick gratings: terminology
  clarification,'' {\em Appl. Optics} {\bf 20}, pp.~3271--3273, 1981.
\doi{10.1364/AO.20.003271}.

\bibitem{Klepp-jpcs16}
J.~Klepp, C.~Pruner, Y.~Tomita, P.~Geltenbort, J.~Kohlbrecher, and M.~Fally,
  ``Advancing data analysis for reflectivity measurements of holographic
  nanocomposite gratings,'' {\em J. Phys.: Conf. Ser.} {\bf 746},
  pp.~012022:1--9, September 2016.
\newblock \doi{10.1088/1742-6596/746/1/012022}.

\bibitem{Fally-oex21}
M.~Fally, Y.~Tomita, A.~Fimia, R.~F. Madrigal, J.~Guo, J.~Kohlbrecher, and
  J.~Klepp, ``Experimental determination of nanocomposite grating structures by
  light- and neutron-diffraction in the multi-wave-coupling regime,'' {\em Opt.
  Express} {\bf 29}, pp.~16153--16163, May 2021.
\doi{10.1364/oe.424233}.

\bibitem{Hadden-PhD24}
E.~Hadden, {\em Polymer based photonic materials for cold neutron optics}.
\newblock PhD thesis, {Universit\"at} Wien, Austria, 2024.
\newblock\doi{10.25365/thesis.77766}

\bibitem{Moharam-josaa95}
M.~G. Moharam, E.~B. Grann, D.~A. Pommet, and T.~K. Gaylord, ``Formulation for
  stable and efficient implementation of the rigorous coupled-wave analysis of
  binary gratings,'' {\em J. Opt. Soc. Am. A} {\bf 12}(5), pp.~1068--1076,
  1995.
\doi{10.1364/JOSAA.12.001068}.

\bibitem{Klepp-ddoi24}
J.~Klepp, M.~Fally, H.~Filter, T.~Jenke, and C.~Pruner, ``Neutron diffraction
  from cyclic allylic sulfide based photopolymer film gratings.''
  \doi{10.5291/ILL-DATA.3-14-445}, 2024.

\bibitem{Drabkin-nima89}
G.~Drabkin, A.~Ioffe, S.~Kirsanov, V.~Zabiyakin, and M.~Bedrizova,
  ``Diffraction-grating interferometer for very cold neutrons: Questions of
  adjustment,'' {\em Nucl. Instrum. Meth. A} {\bf 284}, pp.~176--178, November
  1989.
\newblock \doi{10.1016/0168-9002(89)90274-x}.

\bibitem{Pruner-02}
C.~Pruner, R.~Rupp, M.~Fally, H.~Dachraoui, R.~Mazzucco, J.~Zipfel, and R.~P.
  May, ``Interferometric measurement of the longitudinal coherence-function for
  cold neutrons,'' in {\em {ILL} Annual Report 2002},  G.~Cicognani and
  C.~Vettier, eds., annual report Millennium Programm - New experimental
  techniques, p.~112, ILL, Grenoble, France, 2002.

\end{thebibliography}

\end{document}